\def\pcm3{ {\rm pc}^{-3} }

\def\msun{{\rm M}_{\rm \odot}}

\def\kms{ {\rm km} {\rm s}^{-1}  }

\def\vcirc{v_{\rm circ}}

\def\Mu{M_{\rm U}}
\def\Ml{M_{\rm L}}
\def\Rhalo{R_{\rm halo}}
\def\that{\hat t}
\def\avthat{\langle \hat t \rangle}

\def\tauobsd{\tau_{\rm obsd}}
\def\fBD{f_{\rm BD}}
\def\mi{M_I}
\def\mv{M_V}
\def\RE{R_{\rm E}}
\def\MC{M_{\rm C}}

\def\fr#1#2{\textstyle {#1\over #2}\displaystyle}

\def\spose#1{\hbox to 0pt{#1\hss}}
\def\lta{\mathrel{\spose{\lower 3pt\hbox{$\sim$}}
    \raise 2.0pt\hbox{$<$}}}
\def\gta{\mathrel{\spose{\lower 3pt\hbox{$\sim$}}
    \raise 2.0pt\hbox{$>$}}}

\documentstyle[12pt,aasms4,epsf,rotate]{article}

\begin{document}

\title{Microlensing Halo Models with Abundant Brown Dwarfs}

\author{Eamonn Kerins}
\affil{URA CNRS 1280, Observatoire Astronomique de Strasbourg, 
11 Rue de l'Universit\'e, F-67000 Strasbourg, FRANCE}
\and
\author{N. Wyn Evans}
\affil{Theoretical Physics, Department of Physics, 1 Keble Rd, Oxford,
OX1 3NP, UK}

\begin{abstract} 
All previous attempts to understand the microlensing results towards
the Large Magellanic Cloud (LMC) have assumed homogeneous present day
mass functions (PDMFs) for the lensing populations. Here, we present
an investigation into the microlensing characteristics of haloes with
spatially varying PDMFs and anisotropic velocity dispersion
tensors. One attractive possibility -- suggested by baryonic dark
cluster formation in pregalactic and protogalactic cooling flows -- is
that the inner halo is dominated by stellar mass objects, whereas low
mass brown dwarfs become more prevalent on moving outwards. The
contribution to the microlensing rate must be dominated by dark
remnants ($\sim 0.5\,\msun$) to recover the observed timescales of the
microlensing experiments. But, even though stellar remnants control
the rate, they do not dominate the mass of the baryonic halo, and so
the well-known enrichment and mass budget problems are much less
severe. Using a simple ansatz for the spatial variation of the PDMF,
models are constructed in which the contribution of brown dwarfs to
the mass of the baryonic halo is $\sim 55 \%$ and to the total halo is
$\sim 30 \%$.  An unusual property of the models is that they predict
that the average timescale of events towards M31 is shorter than the
average timescale towards the LMC. This is because the longer line of
sight towards M31 probes more of the far halo where brown dwarfs are
the most common constituent.
\end{abstract}


\section{INTRODUCTION}

Recent results from gravitational microlensing experiments indicate
that at least part of the dynamically-dominant dark Galactic halo has
now been detected~(e.g., Evans 1997). Analyses of the first few years
of data from the MACHO and EROS experiments~(\cite{aubourg93,alcock97})
seemingly show that a substantial fraction of the halo comprises
compact objects which induce microlensing variations on
timescales~\footnote{In this Letter, timescale refers to the Einstein
diameter crossing time} of between 30 and 130 days.  However, the
nature of the lenses remains mysterious. Statistical analyses, which
assume an isothermal halo distribution function populated by objects
with a universal mass function, typically yield lens masses between
$0.1~\msun$ and $1~\msun$ and halo fractions between $20\%$ and
$100\%$, implying a large population of low-mass stars or stellar
remnants~(\cite{alcock97}).  Such inferences contrast sharply with
other observational and theoretical evidence. In particular,
star-count studies undertaken with the Hubble Space Telescope (HST)
place stringent limits on the numbers of low-mass stars and on the age
and spatial density of stellar remnants~(\cite{bahcall,gf,sge}).
Again, the abundance of metals in the interstellar medium limits the
contribution of dark remnants to below that suggested by the
microlensing analyses~(\cite{gm,fms}).  However, this apparent
inconsistency between microlensing on the one hand and deep star-count
and metallicity data on the other may be an artifact of the underlying
assumptions in the adopted models.

In this Letter, we introduce halo models in which key microlensing
parameters -- namely the lens mass, Galactocentric distance and
transverse velocity -- are correlated. How might such correlations
arise? Some of the more promising ``cooling-flow'' theories for the
formation of baryonic dark clusters~(e.g.,
\cite{keith,bernard,depaolis}) predict a spatial gradient in the
present day mass function (PDMF). Here, the inner halo comprises
partly visible stars, which are associated with the spheroid globular
cluster population, whilst the outer halo comprises mostly low-mass
stars and brown dwarfs.  Let us remark that there is evidence for
similar such spatial gradients in the Galactic disc~(\cite{taylor}).
Correlations between the lens velocity and Galactocentric distance are
also possible if the velocity distribution is anisotropic (c.f.,
Markovic \& Sommer-Larsen 1997).  There is strong theoretical
motivation for halo brown dwarfs, both from formation arguments (e.g.,
Ashman 1990, Tegmark et al. 1997) and from constraints on other
candidates (e.g., Carr 1994). But, powerful arguments using the virial
theorem (Gyuk, Evans \& Gates 1998) have shown that the timescales of
the microlensing events preclude the lenses being brown dwarfs. If the
PDMF is the same everywhere, then this also prevents them making a
substantial contribution to the halo.

The aim of this Letter is to show that the microlensing data-set is
consistent with a baryonic component dominated by brown dwarfs. The
crucial point is that, if the PDMF varies with position, the
microlensing rate may be dominated by one mass scale, but the mass
density may be dominated by another one entirely.

\section{CORRELATED HALO MODELS}

This section introduces the haloes under scrutiny. They are
spherically symmetric models which take into consideration the effects
of both varying anisotropy and varying PDMF.  Our starting point is
the rich families of simple power-law densities in power-law
potentials developed by Evans, H\"afner \& de Zeeuw (1997). The
density and rotation curve are
\begin{equation}
\rho \propto r^{-\gamma}, \qquad \vcirc^2 \propto r^{-\beta}.
\end{equation}
Here, we shall choose $\beta =0$ and $\gamma =2$ so that the model is
self-consistent and the rotation curve is asymptotically flat, with
$\vcirc = 220~\kms$.  The velocity dispersions are oriented on a
spherical polar coordinate system and have values
\begin{equation}
\sigma_\phi^2 = \sigma_\theta^2 = (\alpha+1) \sigma_r^2 = {\alpha+1
\over 2\alpha + \beta + \gamma} \vcirc^2.
\end{equation}
When $\alpha =0$, the velocity dispersion tensor is isotropic. When
$\alpha=-1$, the model is composed of radial orbits, while $\alpha
\rightarrow \infty$ corresponds to the circular orbit model. The
distribution of velocities is approximated as a triaxial Gaussian with
these semi-axes.

Let us first consider a halo which is everywhere characterised by a
power-law PDMF, $\phi(M) \propto M^{-n}$, between lower and upper mass
scales $\Ml$ and $\Mu$. In this case, the mass density $\rho \propto
M^{2-n}$ and so is dominated by $\Ml$ if $n > 2$.  By contrast, the
microlensing rate $\Gamma \propto \int \RE \phi(M) dM \propto
M^{3/2-n}$, where $\RE$ is the Einstein radius. This is dominated by
$\Ml$ if $n > \fr32$. Given the assumption that the PDMF is spatially
homogeneous, it follows that if brown dwarfs dominate $\rho$, they
necessarily dominate $\Gamma$. A typical brown dwarf of mass
$0.08\,\msun$ has an Einstein diameter crossing time $\that \sim 30$
days in isothermal models.  The average crossing time $\that$ of the
six-event subsample of Alcock et al. (1997) is $\sim 80$ days. The
absence of events with $\that < 30$ days shows that brown dwarfs
certainly do not dominate the rate and hence the mass density. If,
however, the PDMF is spatially varying, then the above argument no
longer holds, as neither $\that$ nor $\Gamma$ scale simply with $M$
anymore.

Suppose we assume that, at a given Galactocentric radius $r$, the PDMF
is characterised by the simplest of distributions, the delta function:
\begin{equation}
\phi (M) = \frac{\rho}{M} \delta[M-M(r)].
\end{equation}
Motivated by the pregalactic and protogalactic cooling-flow theories
of baryonic dark cluster formation, we take the mass scale $M(r)$ to be
a monotonically decreasing function of $r$ varying like
\begin{equation}
M(r) = \Mu\left( \frac{\Ml}{\Mu} \right)^{r/\Rhalo}, \label{pdmf}
\end{equation}
between the Galactic Center and the halo cutoff radius
$\Rhalo$. Although cooling-flow theories do not specify precisely how
the PDMF varies, the above formula represents one of the simplest and
most convenient ways to parameterise the scenario.  In choosing the
mass scales $\Ml$ and $\Mu$, we assume that the inner halo is
populated mostly by stellar remnants (i.e., $\Mu > 0.8~\msun$, which
is the mass scale at which the stellar main-sequence lifetime equals
the age of the Galaxy). Of course, such mass scales are implicated by
current microlensing results towards the Large Magellanic Cloud
(LMC). The progenitors of these remnants must have been considerably
more massive than $0.8~\msun$.  The outer halo is taken to comprise
almost exclusively brown dwarfs $(10^{-3}~\msun < M < 0.09~\msun)$.
Since eqn~(\ref{pdmf}) implies a smooth transition between the two
mass r\'egimes in the inner and outer halo, there must also be an
intermediate population comprising a mixture of hydrogen-burning stars
and low-mass stellar remnants (which have resulted from progenitors
with masses only a little larger than $0.8~\msun$).  Since we have not
specified the form of the initial mass function, there is some degree
of freedom in this scenario as regards the ratio of remnants to
hydrogen-burning objects in this intermediate regime. Nonetheless, we
will need to consider number-count constraints in some detail in
Section~3.

To determine the consequences for the microlensing observables, let us
examine the set of six models given in Table 1 and see how they fare
in comparison with the six-event sub-sample of the two-year LMC
dataset provided by Alcock et al. (1997). Model~A in Table~1 acts as a
reference, having an isotropic velocity distribution and a homogeneous
PDMF with $M = 0.5~\msun$. It corresponds closely to the best-fit
`standard' model of Alcock et al. (1997). Models B and C characterise
velocity anisotropy and a radially varying PDMF respectively, whilst
Model D combines both attributes. All of the models A--D assume a halo
cutoff radius $\Rhalo$ of 100~kpc, whilst Models E and F incorporate
varying PDMFs within smaller and larger haloes.  For each model, the
overall normalisation is fixed in the following way (Kerins
1998). First, the optical depth contributed solely by the observed
events is
\begin{equation}
\tauobsd = {\pi\over 4 E}\sum_{i=1}^{6} {\that}_i = 5.8 \times 10^{-8}
\end{equation}
where $E$ is the effective exposure ($1.82 \times 10^7$ star
years). This is now used to set the overall normalisation by 
insisting
\begin{equation}
\tauobsd = {\pi\over 4}\int\that \epsilon(\that)
{d \Gamma \over d \that}\,d\that
\end{equation}
where $\epsilon(\that)$ is the detection efficiency of the experiment.
Figure 1 shows the resulting {\em underlying}\/ timescale
distributions for the six halo models A--F. These distributions are
what would be recovered in the limit of perfect detection
efficiency. Their relative normalisation however is fixed by the
observations (via eqns~5 and 6) and thus does take into account the
experimental efficiency.

It is apparent from Figure~1 that the shape of the timescale
distributions is controlled much more by the PDMF and the halo size
than by the velocity distribution. For example, curves~A and B are
very similar, as are curves~C and D. These pairs differ only in the
velocity anisotropy. By contrast, changes in the PDMF can have a
dramatic effect on the shape of the timescale distribution. For
example, curve~C shows an excess of both long and short timescale
events compared to the reference model~A.  The extent of the halo also
has a pronounced effect on the shape of the timescale distribution.
Large haloes, such as model F, show a strong excess towards longer
timescales. This is because most of the low-mass objects now lie
beyond the LMC and so do not contribute at all to the lensing
statistics. In contrast, the small halo model~E peaks sharply at
$\that = 10$~days, with only a modest excess of long-timescale events
relative to reference model~A. This is despite the fact that it has a
somewhat larger upper mass cutoff $\Mu$ than for any of the other
varying PDMF models.  Models~C and D, which correspond to haloes of
intermediate size, show some evidence of a secondary peak at short
timescales ($\that \simeq 10-20$~days). In these two models, the lens
mass drops below $0.5~\msun$ at $r = 22$~kpc, and below $0.09~\msun$
at $r = 43$~kpc -- still somewhat inside the LMC distance. However,
microlensing is relatively insensitive to lenses beyond radii of
$30$~kpc and so their effect on the overall timescale distribution is
slight.

For each of the models, the microlensing observables are reported in
Table~2. The first column gives the number of events that would have
been expected in the two-year LMC data-set. The second column reports
the number of events with crossing times less than 20 days. Alcock et
al. (1997) found no events with such short crossing times, implying a
$1\, \sigma$ upper limit of 1.1 events on the underlying
distribution. Comparison with the observed quantities in the last row
of Table~2 indicates that all but model E reproduce the observed
number of events within the uncertainties. Model E is marginally
excluded at the $1 \,\sigma$ level by the upper limit on the number of
events with $\that < 20$~days.  Column~3 gives the mean Einstein
crossing time, after accounting for detection efficiencies. All models
with the exception of model F lie within the $1 \, \sigma$ errors.
Column~4 gives the total optical depth $\tau$. This is slightly
different for each of the models, because it is only the observed
optical depth $\tauobsd$ that is fixed by the normalisation.  The
remaining two columns give the baryonic mass fraction $f$ and the
brown dwarf mass fraction $\fBD$. These have been computed assuming a
local halo density of $\rho_0 = 0.01 \msun \pcm3$
(c.f. \cite{bernard,ggt}). The most important conclusion to be drawn
from the Table is that strongly inhomogeneous mass functions can give
substantial boosts to the mass fraction of brown dwarfs. For example
in model~C, which is comfortably within the $1 \, \sigma$ error limits
for all the microlensing observables, brown dwarfs provide $56\%$ of
the mass of the baryonic halo. The contribution of brown dwarfs to the
overall halo mass is also significant ($\sim 29\%)$.

\section{TESTS AND PREDICTIONS}

This is a provocative model and it is susceptible to observational
checks and constraints. An obvious concern is whether it violates the
limits on hydrogen-burning stars. Flynn, Gould \& Bahcall (1996) have
used the Hubble Deep Field ($\ell = 126^\circ, b = 55^\circ$,
hereafter HDF) -- the deepest optical field ever obtained -- to set
constraints on the baryonic fraction of the halo. These constraints
amount to an upper limit of about $1\%$ on the halo fraction in
hydrogen burning objects if they are unclustered.  In the scenario
which motivates the present work, the stars are predicted to be
clustered, which therefore modifies the number count constraints
somewhat (c.f., Kerins 1997).  Some of the stars have masses well
above the hydrogen-burning limit, but are confined spatially to shells
whose radius depends on the mass of the star through eqn~(4). As well
as constraints from `pencil beam' fields such as HDF, it is also
important to verify that the models are consistent with the
microlensing experiments themselves, which do not observe a
significant stellar population in front of the LMC.  To this end, we
shall compute expected cluster number counts for both the HDF $I$ band
and MACHO $V$ band by employing the photometric predictions of Table~4
of D'Antona \& Mazzitelli (1996). These apply to stellar populations
of age 10~Gyr and metallicity $Z = 4 \times 10^{-3}$. The $V$ and $I$
band predictions are well fit by the following fourth-order polynomial
least-squares fits:
\begin{eqnarray}
{M\over \msun} & = & -3.16 \times 10^{-4} \, \mi^4 + 0.0125 \,
\mi^3 - 0.168 \, \mi^2 + 0.809 \, \mi - 0.469 \nonumber \\
& = & -1.20 \times 10^{-4} \mv^4 + 5.92 \times 10^{-3}
\mv^3 - 0.0987 \, \mv^2 + 0.580 \, \mv - 0.313.
\end{eqnarray}
These fits are valid between $0.75 < V-I < 4.50$, corresponding to
$0.80\msun > M > 0.09\msun$. Since we are only interested in the halo
models with varying PDMFs, we confine our attention to models~C, E and
F (of course, the predictions for model~D are identical to those of
model~C).

Taking the HDF field of view as 4.4~arcmin$^2$, using the HDF point
source magnitude limits $24.63 < I < 26.3$ as calculated by Flynn,
Gould \& Bahcall (1996), and assuming that all cluster stars can be
resolved, we find that the expected cluster number counts for models
(C,E,F) are $N_{\rm C} \sim (5.2,2.3,9.6)\times 10^{-3} \,
(f/0.5)(10^5 \msun/\MC)$, where $\MC$ is the cluster mass. Thus, for
clusters between $10^4 - 10^5~\msun$, all models are comfortably
within the Poisson $95\%$ confidence upper limit of 3 clusters based
on no detections. In the case of MACHO, assuming a field of view of
11~deg$^2$ and a limiting $V$-band magnitude of 21, we obtain $N_{\rm
C} \sim (0,1.9,0) \times (f/0.5)(10^5 \msun/\MC)$. This is again
within the $95\%$ upper limit for a null detection, though model E
requires $M_{\rm C} > 6\times 10^4~\msun \, (f/0.5)^{-1}$.  Haloes
with a cut-off radius smaller than 50~kpc produce too many clusters
within the MACHO field of view. The HDF is mostly sensitive to the
numerous low-mass stars further out in the halo and does not sample
the nearer, more massive stars as well. As a result, it does not
seriously constrain any of the models. For models C and F, the MACHO
sensitivity is insufficient to detect any hydrogen burning stars, the
nearest of which lie about 16~kpc away for model~C and 24~kpc away for
model~F.  Locally, the halo density is dominated by stellar remnants
for the spatially varying PDMF models, so one may imagine constraints
from white-dwarf number counts (c.f. \cite{glf}). However, if
clustered, the nearest remnants are several kpc from us, and so too
faint to detect.


There are some obvious tests of this scenario. First, the timescale
distributions are broader in models with strong inhomogeneities in the
mass function. This is evident from Figure~1, where models~C to F have
timescale distributions with long tails and consequently larger second
moments than the models with weak or no inhomogeneities.  This test
may be difficult to perform until a much larger number of events are
recorded.  A better prospect may be to combine information along
differing lines of sight. A distinguishing characteristic of the
models dominated by brown dwarfs in the outer parts concerns the ratio
of the average Einstein diameter crossing time $\avthat$ towards the
LMC, SMC and M31. The ratios for our reference model A are $1 : 1 :
1.6$ (LMC:SMC:M31), whereas in model C they are $1 :1.1 :0.6$.  This
effect does not hold for model~E because it requires $\Rhalo$ to be
larger than the distance to the LMC and SMC. It holds for models C, D
and F because in these models there are many brown dwarfs which are
only probed by the longer lines of sight to M31.  It is very hard to
see how this effect can be produced in any other way, so this is a
definitive test for such models.

Are there any possible tests of this scenario as applied to external
galaxies?  Sackett et al. (1994) discovered a faint, luminous halo
surrounding the spiral galaxy NGC 5907 and suggested that it might be
composed of faint M dwarfs or possibly brown dwarfs. Now, brown dwarfs
are brightest at mid-infrared wavelengths.  This led Gilmore \&
Unavane (1998) to examine 4 edge-on spiral galaxies with the Infrared
Space Observatory (ISO) camera. They concluded that low mass stars or
young brown dwarfs do not make a significant contribution to the dark
haloes of at least two of the galaxies (NGC 2915 and UGC 1459), though
brown dwarfs older than 1 Gyr remain viable. As Gilmore \& Unavane
(1998) point out, once the origin of noise in the ISO data is better
understood, these conclusions may become stronger. The next generation
of space-based infrared observatories will provide stringent checks on
the contribution of baryonic dark clusters to the haloes of external
galaxies. This remains a very promising way to confirm or confute this
theory.

\section{CONCLUSIONS}

The assumption of a uniform present day mass function (PDMF)
throughout the Galactic halo has been made in all previous studies of
the microlensing dataset. Of course, it is natural enough to make the
simplest assumption, but there is no evidence to suggest that it is
correct. The main thrust of this Letter is to point out that the
impasse of the microlensing results can be overcome by discarding this
unwarranted assumption.

Although our own PDMF is not likely to be correct in detail, its
general form does receive physical support from baryonic dark cluster
formation theories~(e.g., \cite{keith,bernard,depaolis}). These
suggest that there may be a gradation of masses of baryonic objects in
the halo, with larger mass objects in the inner regions and abundant
brown dwarfs in the far halo.  Such models can reproduce the
microlensing observables within the uncertainties. Such models can be
made consistent with Hubble Space Telescope limits on faints stars.
The important point is that in models with spatially varying mass
functions, {\it brown dwarfs need not dominate the microlensing rate
but they can still dominate the mass of the baryonic halo}. In such
models, the lenses are predominantly dark stellar remnants, but these
now comprise a much smaller fraction of the halo. So, this evades the
embarrassing mass budget and the chemical enrichment problems that
occur when the halo has a uniform PDMF~(\cite{gm,fms}).

\acknowledgments
EK is supported by an EC Marie Curie Training and Mobility of
Researchers Fellowship and thanks the sub-Department of Theoretical
Physics, Oxford for hospitality.  NWE is supported by the Royal
Society and thanks Gerry Gilmore for providing material in advance
of publication.

\eject

\begin{deluxetable}{cccccl}
\tablecaption{Catalogue of Correlated Halo Models}
\tablehead{
\colhead{Model} & 
\colhead{$\alpha$} &
\colhead{$\Ml/\msun$} &
\colhead{$\Mu/\msun$} &
\colhead{$\Rhalo$/kpc} & 
\colhead{Description}
}
\startdata
A & 0 & 0.5 & 0.5 & 100 & Reference model: homog. PDMF \nl
B & -0.5 & 0.5 & 0.5 & 100 & Homog. PDMF, anisotropic vel. \nl
C & 0 & $10^{-3}$ & 3 & 100 & Varying PDMF, isotropic vel. \nl
D & -0.5 & $10^{-3}$ & 3 & 100 & Varying PDMF, anisotropic vel. \nl
E & 0 & $10^{-3}$ & 10 & 50 & Varying PDMF, small halo \nl
F & 0 & $10^{-3}$ & 2 & 200 & Varying PDMF, large halo \nl
\enddata
\end{deluxetable}

\eject

\begin{deluxetable}{ccccccc}
\tablecaption{Microlensing Predictions of the Halo Models for the MACHO 2-year 
LMC Dataset. The last row shows the current observational status,
together with 1-$\sigma$ errors on the mean values.}
\tablehead{
\colhead{Model} & 
\colhead{$N_{\rm obs}$} &
\colhead{$N_{\rm obs}(\hat{t} < 20$~d)} &
\colhead{$\langle \hat{t} \rangle$/d } &
\colhead{$\tau/10^{-7}$} & 
\colhead{$f$} &
\colhead{$f_{\rm BD}$}
}
\startdata
A & 5.5 & 0.06 & 87 & 2.23 & 0.48 & 0 \nl
B & 5.9 & 0.14 & 81 & 2.29 & 0.49 & 0 \nl
C & 4.6 & 0.06 & 104 & 2.40 & 0.52 & 0.29 \nl
D & 4.9 & 0.07 & 98 & 2.42 & 0.52 & 0.29 \nl
E & 6.3 & 1.16 & 76 & 2.42 & 0.52 & 0.25 \nl
F & 4.1 & 0.01 & 118 & 2.53 & 0.54 & 0.32 \nl
Obsd & $6\pm 2.4$ & $0^{+1.1}_{-0}$ & $80\pm 35$\tablenotemark{a}
 & $>2.06^{+1.12}_{-0.73}$ & $>0.44^{+0.24}_{-0.16}$ & \nodata \nl
\tablenotetext{a} {Using blend-corrected timescales}
\enddata
\end{deluxetable}

\eject

\begin{figure}
\rotate[r]{
               \epsfxsize 0.75\hsize 
               \epsffile{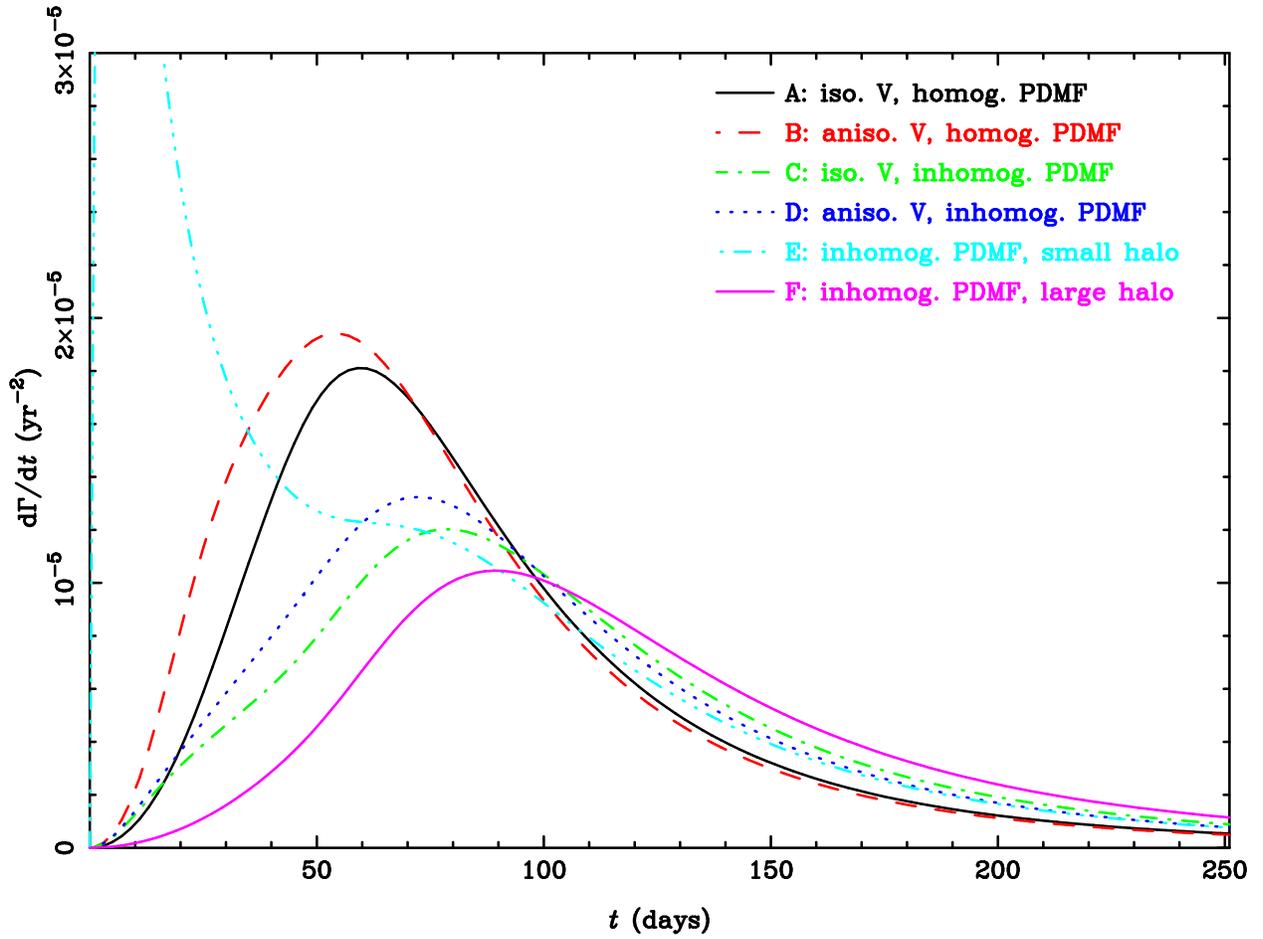}
}
\caption{This figure shows the timescale distributions for the
six halo models `A' to `F'. The curves have been normalised so
that they reproduce the optical depth in the observed events
(see text). 
 }
\end{figure}

\end{document}